\documentclass[%
 reprint,
superscriptaddress,
 amsmath,amssymb,
 aps,prl
]{revtex4-1}

\usepackage{graphicx}
\usepackage{dcolumn}
\usepackage{bm}
\usepackage{amsmath}
\usepackage{amssymb}

\usepackage{hyperref}
\usepackage[dvipsnames]{xcolor}
\hypersetup{
	colorlinks,
	linkcolor={red!80!black},
	citecolor={Blue},
	urlcolor={MidnightBlue}
}

\begin{document}
\title{Reconfigurable electro-optic frequency shifter}

\author{Yaowen Hu}\affiliation{John A. Paulson School of Engineering and Applied Sciences, Harvard University, Cambridge, MA 02138, USA}\affiliation{Department of Physics, Harvard University, Cambridge, MA 02138, USA}

\author{Mengjie Yu}
\author{Di Zhu}\affiliation{John A. Paulson School of Engineering and Applied Sciences, Harvard University, Cambridge, MA 02138, USA}

\author{Neil Sinclair}\affiliation{John A. Paulson School of Engineering and Applied Sciences, Harvard University, Cambridge, MA 02138, USA}
\affiliation{Division of Physics, Mathematics and Astronomy, and Alliance for Quantum Technologies (AQT), California Institute of Technology, 1200 E. California Boulevard, Pasadena, CA 91125, USA}

\author{Amirhassan Shams-Ansari}
\author{Linbo Shao}
\author{Jeffrey Holzgrafe}
\author{Eric Puma}\affiliation{John A. Paulson School of Engineering and Applied Sciences, Harvard University, Cambridge, MA 02138, USA}

\author{Mian Zhang}\affiliation{HyperLight Corporation, 501 Massachusetts Ave, Cambridge, MA 02139, USA
}

\author{Marko Loncar}\email{loncar@seas.harvard.edu}
\affiliation{John A. Paulson School of Engineering and Applied Sciences, Harvard University, Cambridge, MA 02138, USA}

\begin{abstract}
Efficient and precise control of the frequency of light on gigahertz scales is important for a wide range of applications. Examples include frequency shifting for atomic physics experiments~\cite{Greiner2002,Hu2017}, single-sideband modulation for microwave photonics applications~\cite{Supradeepa2012,Marpaung2019,Fandino2016,Eggleton2019}, channel switching and swapping in optical communication systems~\cite{Yoo1996,Lukens2020}, and frequency shifting and beam splitting for frequency domain photonic quantum computing~\cite{Kues2017,Kobayashi2016,Lukens2016,Lu2018,Joshi2020}. However, realizing GHz-scale frequency shifts with high efficiency, low loss and reconfigurability, in particular using a miniature and scalable device, is challenging since it requires efficient and controllable nonlinear optical processes. Existing approaches based on acousto-optics~\cite{Eggleton2019,Kittlaus2018,Sohn2018,Liu2019}, all-optical wave mixing~\cite{Kobayashi2016,Joshi2020,Huang1992,Li2016,Heuck2019}, and electro-optics~\cite{Preble2007,Johnson1988,Izutsu1981,Wright2017} are either limited to low efficiencies or frequencies, or are bulky, and have yet to simultaneously demonstrate the required properties mentioned above. Here we demonstrate an on-chip electro-optic frequency shifter that is precisely controlled using only a single-tone microwave signal. This is accomplished by engineering the density of states of, and coupling between, optical modes in ultra-low loss electro-optic waveguides and resonators realized in lithium niobate nanophotonics~\cite{Zhang2017}. Our device provides frequency shifts as high as 28 GHz with measured shift efficiencies of $\sim$99\% and insertion loss of $<$0.5 dB. Importantly, the device can be reconfigured as a tunable frequency-domain beam splitter, in which the splitting ratio and splitting frequency are controlled by microwave power and frequency, respectively. Using the device, we also demonstrate (non-blocking) frequency routing through an efficient exchange of information between two distinct frequency channels, i.e. swap operation. Finally, we show that our scheme can be scaled to achieve cascaded frequency shifts beyond 100 GHz. Our device could become an essential building-block for future high-speed and large-scale classical information processors~\cite{Yoo1996,Frankel1998} as well as emerging frequency-domain photonic quantum computers~\cite{Kues2017,Lukens2016}.
\end{abstract}

\maketitle

The progress of photonic science and technology is intimately related to the ability to utilize and precisely control all fundamental degrees of freedom of a photon. While, for example, photon position (path) and polarization can be readily controlled using elementary optical components such as beam splitters and waveplates, frequency control is more challenging as it requires changing the energy of a photon. Typical devices used for frequency control of light are based on acousto-optics, all-optical wave-mixing, and electro-optics (EO). Acousto-optic modulators use phonon scattering to control photon energy and can shift the frequency of light in the kHz to few GHz range~\cite{Kittlaus2018,Sohn2018}. However, they have demonstrated high efficiency only in bulk structures that are not compatible with photonic integration. All-optical wave-mixing can achieve efficient frequency conversion in the THz range in bulk media~\cite{Kobayashi2016,Joshi2020,Huang1992}. However, it requires stringent phase matching conditions, can suffer from parasitic nonlinear processes, and is difficult to control due to a nonlinear dependence on optical power. 

The EO effect, utilized in modulators that are widely used in modern telecommunication networks, directly mixes microwave and optical fields and can be used to achieve frequency control of light. However, traditional EO modulators inevitably produce undesired symmetric sidebands and are thus unable to achieve efficient frequency shifts. A number of methods have been developed to circumvent this problem. Serrodyne modulation uses a saw-tooth waveform to generate unidirectional frequency shifts~\cite{Johnson1988}, but extending this method to the GHz regime requires broadband and high-power electronics, which ultimately limits its practical usage. In-phase and quadrature (IQ) modulators can eliminate symmetric sidebands via destructive interference among multiple modulators~\cite{Izutsu1981}, but they are fundamentally limited by biasing-induced insertion loss and are accompanied by higher-order sidebands. Other methods such as adiabatic tuning of the optical cavity resonance~\cite{Preble2007} or spectral shearing~\cite{Wright2017,Fan2016}, which applies a linear temporal phase to light with sinusoidal modulation, are capable of unidirectional frequency shifting, but they require pulsed operation with a known timing reference. Consequently, a practical and efficient EO device that can shift the frequency of light on-demand is still missing.

\begin{figure*}
\includegraphics[width = 5.3in]{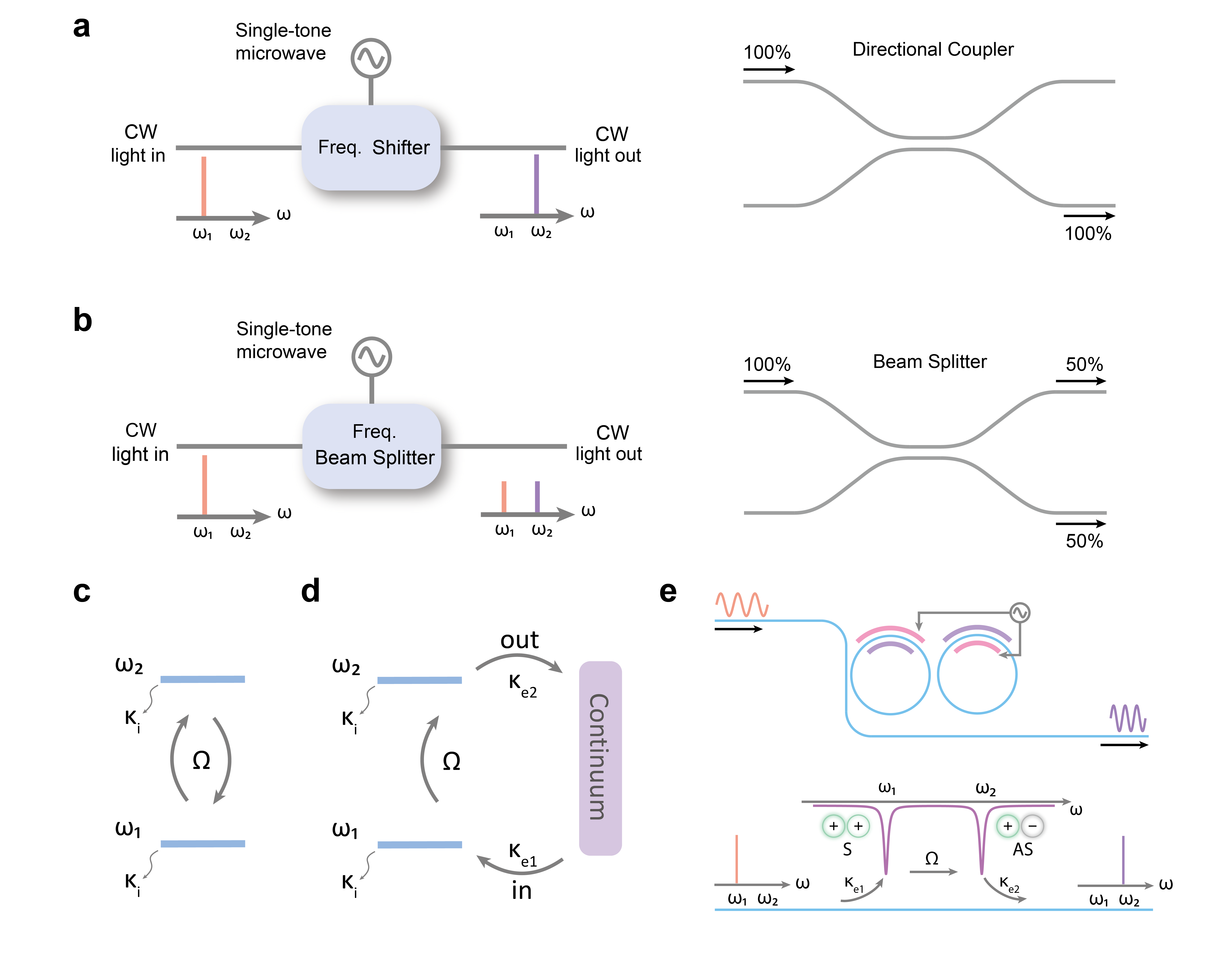}
\caption{\label{fig1} Concept of the reconfigurable electro-optic frequency shifter. a, A frequency shifter converts light from one frequency to another, analogous to a directional coupler that changes the path (spatial modes) of a photon. b, A frequency beam splitter partially converts light from one frequency to another, similar to its spatial-mode counterpart. c, d, Principle of a generalized critical coupling condition that allows complete transfer of energy between levels vis-a-vis frequency shifting of light. c, A coherent coupling $\Omega$ causes oscillation of energy between two discrete levels $\omega_1$ and $\omega_2$ (e.g. two cavity modes), each has a decay rate of $\kappa_i$. d, Frequency shifting by coupling two discrete levels to a continuum. Light is coupled from the continuum to level $\omega_1$ at a rate $\kappa_{e1}$, from level $\omega_1$ to level $\omega_2$ at a rate $\Omega$, and from level $\omega_2$ to the continuum at a rate $\kappa_{e2}$. The critical coupling condition $\kappa_{e1}=\Omega=\kappa_{e2}\gg\kappa_i$ results in a unidirectional flow of energy. e, Schematic of the device used to implement the generalized critical coupling condition (top), and its frequency-domain representation (bottom). The coupled-ring system provides a pair of hybrid modes, referred to as symmetric (S) and anti-symmetric (AS). Coupling between them is induced by electro-optic modulation.}
\end{figure*}

Here we overcome these limitations and demonstrate an on-chip EO frequency shifter that has $\sim$99\% shift efficiency and low insertion loss. Importantly, this is accomplished using only a single monotone microwave source. Our EO frequency shifter acts only on selected frequency modes without affecting other frequencies of light. Furthermore, it features a tunable shift efficiency in the 0\%-99\% range, that can be controlled by the applied microwave power. At maximum efficiency, the frequency of all inserted photons is shifted to another frequency. In this regime the device is analogous to a directional coupler (Fig.~\ref{fig1}a): it swaps two modes, but in the frequency rather than the spatial domain. On the other hand, at e.g. 50\% shift efficiency, the device serves as a 50:50 frequency domain beam splitter (Fig.~\ref{fig1}b). By changing the power of the microwave signal, the splitting ratio can be controlled, and tunable frequency domain beam splitter can be realized. These operations represent the fundamental functionalities required for controlling the frequency degree of freedom of a photon, in analogy with the control of its polarization and path.

\begin{figure*}
\includegraphics{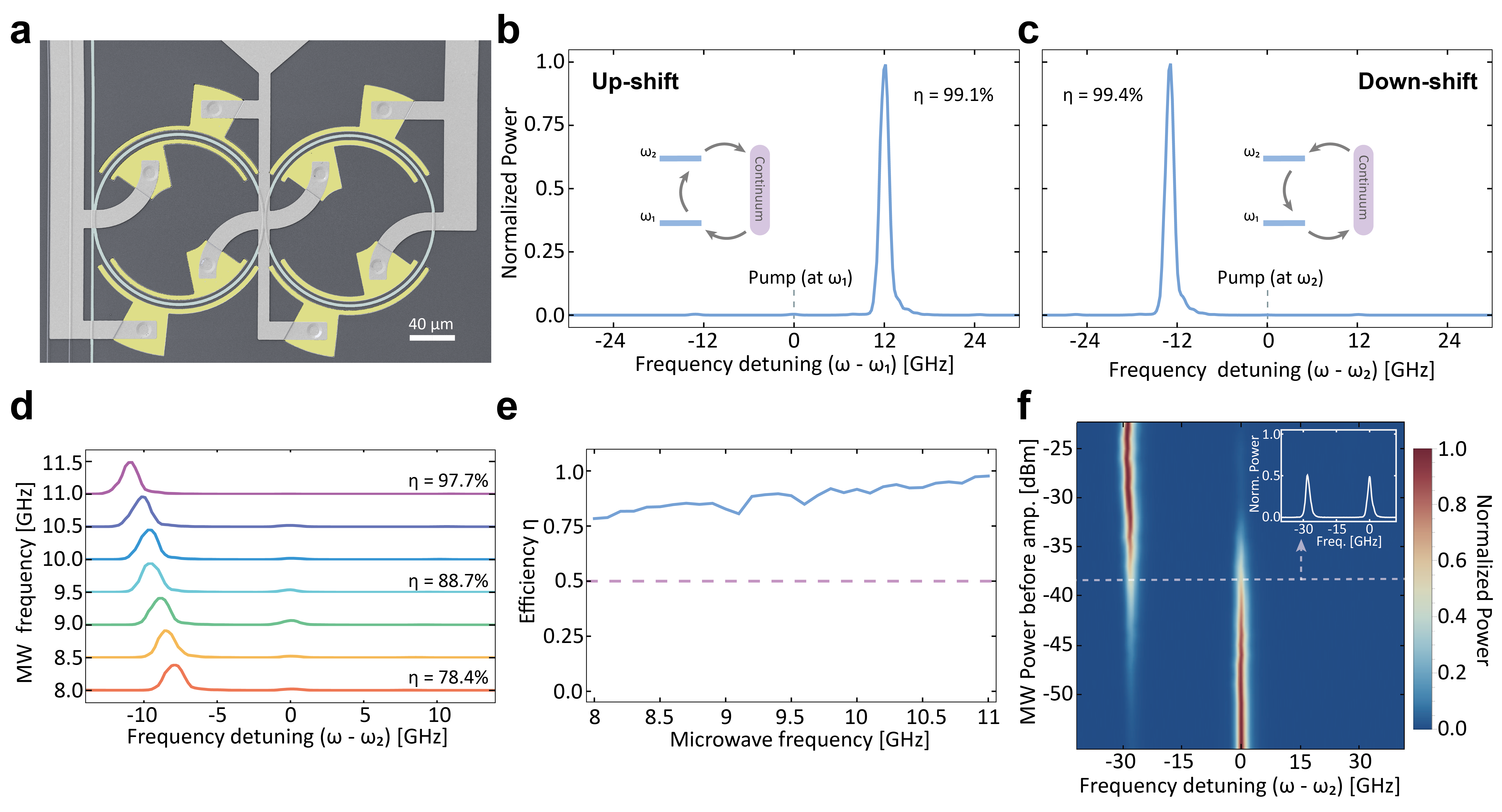}
\caption{\label{fig2} Reconfigurable electro-optic frequency shifter. a, Scanning electron microscopy (SEM) micrograph of the frequency shifting device in false color. The light blue represents the optical waveguide and ring resonators. Metal electrodes have two layers, connected by vias through the cladding oxide. Top and bottom electrode layers are represented by light grey and yellow, respectively. The electrodes are designed to minimize parasitic capacitance and inductance to achieve efficient modulation at high microwave frequencies. b, c, Electro-optic frequency shifter. Up- (b) and down- (c) shifts of 12.5 GHz with $>$99\% shift efficiency at telecommunication wavelengths. The insets show the directions of energy flow in the energy-level description of Fig.~\ref{fig1}d. d, Frequency shift with microwave detuning. The efficiency is reduced from near-unity to 79.2\% as the microwave frequency is detuned from 11 GHz (equal to the doublet splitting) to 8 GHz. e, Shift efficiency for varied microwave detuning. The frequency range is limited by our microwave amplifier. f, Tunable frequency beam splitter. Increasing the microwave power (Y-axis, before amplifier) changes the shift efficiency continuously from 0 to near-unity at a shift-frequency of 28.2 GHz, allowing for a frequency beam splitter with a tunable splitting ratio. Inset: output optical spectrum at 50:50 splitting. MW, microwave; Amp, amplification; Norm, normalized.}
\end{figure*}

To realize the frequency shifting functionality, we introduce a general method to control the flow of light in the frequency domain. We consider two discrete photonic energy levels (Fig.~\ref{fig1}c), which could be two resonances of an optical cavity or a doublet formed by mode anti-crossing. When driven, e.g. using coherent microwave signals and electro-optic effect considered in this work, such a two-level photonic system undergoes Rabi oscillation with rate $\Omega$. As a result, the frequency of light inside such a system oscillates between two levels. To enable an efficient unidirectional frequency shift, we introduce a continuum of levels (e.g. optical waveguide) that couples to both discrete levels. Importantly, by controlling the coupling rates between different levels the photons can be injected at one discrete level and extracted from the other one (Fig.~\ref{fig1}d). Photons of frequency $\omega_1$ are coupled from the continuum into level $\omega_1$ with a rate of $\kappa_{e1}$, while being coupled out of level $\omega_2$ back to the continuum with a rate $\kappa_{e2}$. Assuming that each of the levels has a negligible intrinsic loss rate $\kappa_i\ll\kappa_{e1}, \kappa_{e2}$, complete energy transfer (100\% frequency shift) occurs when these three rates of energy exchange are balanced: $\kappa_{e1}=\Omega=\kappa_{e2}$ (see Supplementary Materials for a detailed discussion on this condition). We refer to this as a generalized critical coupling condition. Otherwise, when the condition is not met, partial energy transfer occurs between the two levels, and some fraction of the photons at frequency $\omega_1$ are coupled back into the continuum. This is used to realize a tunable frequency domain beam splitter. Notably, when the critical coupling condition is satisfied, the reverse frequency conversion process will also be in a balanced state: photons injected in the continuum at frequency $\omega_2$ will be converted to $\omega_1$ and outcoupled to the continuum at that frequency.

\begin{figure*}
\includegraphics{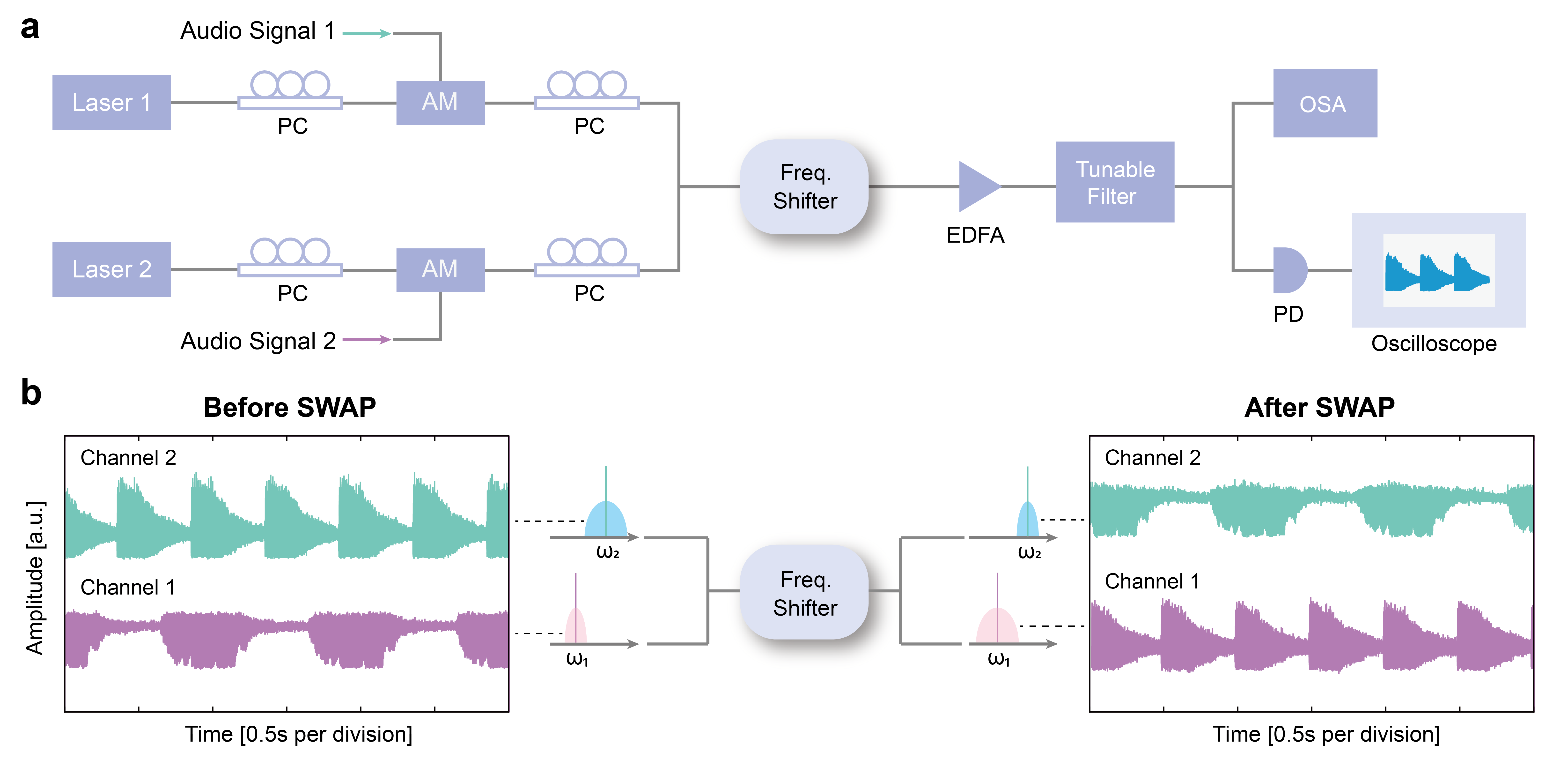}
\caption{\label{fig3} Information swapping between two frequency channels. a, Experimental setup. Laser 1 and 2 emit light of frequency $\omega_1$ and $\omega_2$, respectively, with 11 GHz detuning. The two beams are independently modulated to carry unique audio signals, combined, and then sent to the frequency shifter. The output optical spectrum is monitored by an optical spectrum analyzer, and each of the frequency channels is measured using a photodetector followed by a real-time oscilloscope after passing a tunable optical filter. AM, amplitude modulator; PC, polarization controller; OSA, optical spectrum analyzer; EDFA, erbium-doped fiber amplifier; PD, photodetector. b, Time-domain audio signal in each channel before and after the swapping. The purple and green traces represent the audio signal in channel 1 ($\omega_1$) and channel 2 ($\omega_2$), respectively. }
\end{figure*}

We experimentally implement the proposed scheme using a coupled-cavity structure on thin-film lithium niobate (Fig.~\ref{fig1}e), in which coherent coupling of two optical frequency modes is achieved via the EO effect and microwave driving~\cite{Zhang2018,Wade2015}. Evanescent coupling between two identical ring resonators gives rise to a resonance doublet that corresponds to symmetric (S) and anti-symmetric (AS) modes with frequencies $\omega_1$ and $\omega_2$. A single bus waveguide provides a continuum of modes as well as the input and output ports of our device. The two cavities are efficiently modulated using a single sinusoidal microwave drive (Fig.~\ref{fig1}e, top), and support high frequencies owing to the small capacitance and parasitic inductance of the electrodes (Fig.~\ref{fig2}a and Supplementary Materials). The microwave frequency is either matched or detuned from the frequency difference between the S and AS modes of the doublet, depending on the experiment we perform. The coupling rate $\gamma$ between the waveguide and the cavity is 30 times higher than the intrinsic loss $\kappa_i$ of the cavity, yielding two strongly over-coupled modes with balanced effective mode-waveguide coupling of $\kappa_{e1}=\kappa_{e2}=\gamma/2$ that are needed for the generalized critical coupling condition (see Supplementary Materials). Specifically, we fabricate devices with various doublet splittings of 11.0, 12.5 and 28.2 GHz at telecommunication wavelengths by changing the gap between the coupled rings.

We first demonstrate the frequency shifting feature of our device. Continuous-wave (CW) light of frequency $\omega_1$ (wavelength of 1601.2 nm) and a microwave tone of frequency 12.5 GHz, which matches the splitting of the resonance doublet, are sent to the device. The continuous optical frequency spectrum at the output of the device (Fig.~\ref{fig2}b) shows that nearly all the power at frequency $\omega_1$ (S mode) is converted to frequency $\omega_2$ (AS mode), with a measured shift efficiency of 99.1\%. This efficiency is defined as the ratio of the optical powers of the shifted frequency and the total light output: $\eta=P_\mathrm{shift}/P_\mathrm{out}$. Importantly, our device also operates in reverse: pumping at frequency $\omega_2$ (AS mode) leads to a down-shift to frequency $\omega_1$ with a measured  $\eta$=99.4\% (Fig.~\ref{fig2}c). The device has a low on-chip insertion loss $IL$ of only 1.2 dB (see Supplementary Materials), defined as the ratio of output and input powers, $IL=P_\mathrm{out}/P_\mathrm{in}$. The amount of energy that is transferred to the target frequency is captured by $\eta$, while $IL$ represents the optical loss due to the propagation of light in the device. The absolute on-chip efficiency of our device is therefore $\eta\times IL=P_\mathrm{shift}/P_\mathrm{in}$. 

Second, we show that frequency shifts over a large microwave bandwidth can be achieved by tuning the microwave frequency. Light is injected into the AS mode (at frequency $\omega_2$) of a device with an 11.0 GHz doublet splitting to realize a frequency down-shift. As we vary the microwave frequency from 11.0 GHz to 8.0 GHz, $\eta$ reduces from 97.7\% to 78.4\%, indicating a 3 dB bandwidth of $>$3 GHz that is currently limited by the bandwidth of our microwave amplifier (Fig.~\ref{fig2}d and e). The bandwidth of the shifter benefits from the strong over-coupling of the optical resonators to the optical waveguide and strong microwave modulation, yielding a bandwidth beyond the cavity linewidth ($2\pi\times2.8$ GHz, see Supplementary Materials). One advantage of our approach is that increasing the doublet splitting for larger frequency shift does not degrade $\eta$ or $IL$, as long as the generalized critical coupling condition is satisfied. This is experimentally confirmed using a 28.2 GHz device with a measured $\eta$ = 98.7\% and $IL$ = 0.45\,dB. 

\begin{figure*}
\includegraphics{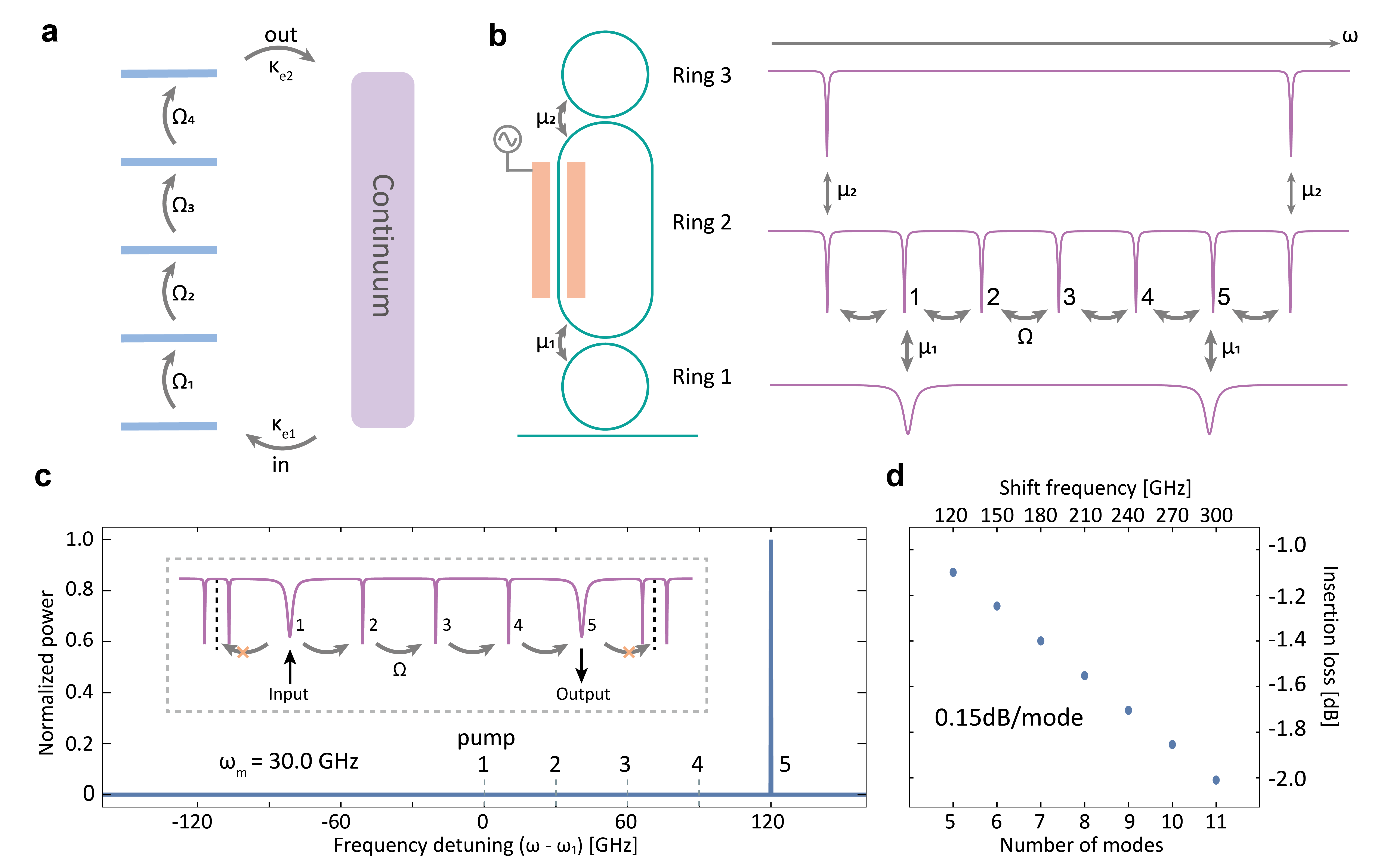}
\caption{\label{fig4} Cascaded frequency shifting. a, Balancing of coupling rates between several energy levels can permit unidirectional energy flow when $\kappa_{e1}=\Omega_1=\Omega_2=\cdots=\kappa_{e2}$. b, Proposed structure to realize cascaded frequency shifting. It consists of three coupled rings: ring 2 provides a set of equally detuned energy levels, with a mode coupling rate of $\Omega$ induced by microwave modulation; ring 1 over-couples modes 1 and 5 to the waveguide; and ring 3 induces a boundary to the cascade process by mode splitting. $\mu_1$: coupling rate between ring 1 and 2; $\mu_2$: coupling rate between ring 2 and 3. c, Simulated optical spectrum shows cascaded frequency shift when the generalized critical coupling condition is satisfied. A single microwave tone at 30 GHz (equal to the FSR of ring 2) generates a 5-mode cascaded frequency shift of 120 GHz with an insertion loss of 1.1 dB. The inset illustrates the energy flow in the coupled cavities system. $\omega_1$ in the x-axis is the frequency of mode 1 in the inset. d, The shift can be scaled to a larger number of modes, with an incremental insertion loss of 0.15 dB per mode. }
\end{figure*}

Next, we reconfigure the 28.2 GHz device into a frequency beam splitter with a fully tunable power splitting ratio by varying the power of the microwave signal. As the microwave power is increased, $\eta$ of a down-shift continuously increases from 0 to near-unity (Fig.~\ref{fig2}f). The case in which optical power is split equally between the two frequency modes $\omega_1$ and $\omega_2$ yields a 50:50 frequency beam splitter (as illustrated in Fig.~\ref{fig1}b and with results shown in the inset of Fig.~\ref{fig2}f). It should be noted that the frequency splitter, which is based on coherent mixing of optical and microwave fields, performs unitary operations (in the limit of vanishingly small loss) identical to a standard 4-port configuration of a tunable spatial-mode beam splitter (see Supplementary Materials for theoretical analysis).

To show that our device can perform up- and down- shifts simultaneously, we demonstrate information swapping between two different frequency channels using a single device, mimicking a 4-port coupler, as depicted in Fig.~\ref{fig1}a. Such an operation allows the exchange of information between frequency channels without detection, which is of particular importance for modern telecommunications and quantum information processing. The measurement setup is shown in Fig.~\ref{fig3}a. Two laser beams of frequency $\omega_1$ and $\omega_2$ are separately modulated to carry two different audio signals. The frequency difference between $\omega_1$ and $\omega_2$ is set to be 11 GHz, matching the doublet splitting of the device. The two laser beams are then combined and sent into our device. At the output, we use a tunable filter to select each frequency channel and measure its time-domain audio signal using a photodetector followed by a real-time oscilloscope. Figure~\ref{fig3}b shows the measured audio signals before and after the swap operation, which exhibits little distortion owing to the near-unity shift efficiency. 

Finally, we show that our concept can be generalized to shift optical frequencies beyond 100 GHz. Although this can be achieved using devices we discussed so far, by further increasing the evanescent coupling between the resonators (for larger doublet splitting) and by using high-frequency microwave amplifiers, this approach will ultimately be limited by the RC limit of the electrode and cost of high-speed microwave electronics. To mitigate this, we propose a generalized cascaded shifter by extending our scheme to a system consisting of multiple discrete levels coupled to a continuum (Fig.~\ref{fig4}a). When coherent coupling applied between nearest-energy levels satisfies the generalized critical coupling condition $\kappa_{e1}=\Omega_1=\Omega_2=\cdots=\kappa_{e2}$, the energy can flow unidirectionally. As a result, a complete frequency shift can be achieved. Figure~\ref{fig4}b depicts a system that can realize this process, comprised of only three resonators and a single bus waveguide. A larger microresonator (ring 2) provides a family of discrete modes separated by the free spectral range (FSR, e.g. 30 GHz). Modulating this resonator at a frequency equal to the FSR provides coherent coupling between nearest-neighbor frequency modes~\cite{Yuan2018,Reimer2019}. An over-coupled smaller microresonator (ring 1) with a large FSR (e.g. 120 GHz) is used to over-couple two modes (e.g. mode 1 and mode 5 in Fig.~\ref{fig4}b) of ring 2 to the bus waveguide. Another microresonator (ring 3, FSR=180 GHz) induces mode crossing and thus breaks the cascade process in ring 2 since the modulation no longer couples the resultant doublet levels with their nearest-neighbor levels. The inset of Fig.~\ref{fig4}c illustrates the resonance spectrum of this device, which permits unidirectional energy flow under the generalized critical coupling condition. The main panel of Fig.~\ref{fig4}c shows that the simulated shift efficiency can reach near-unity with a 1.1 dB insertion loss for a 120 GHz shift using a single 30 GHz microwave drive. Larger frequency shifts $(n-1)\omega_m$ can be realized by increasing the number of modes $n$ without compromising the high shift efficiency, at a cost of an added insertion loss of 0.15 dB (simulated) per mode (Fig.~\ref{fig4}d). Details of the simulations are available in Supplementary Materials.

In summary, we proposed and demonstrated an efficient, low-loss, and reconfigurable electro-optic frequency shifter, enabled by recent breakthroughs in integrated lithium niobate photonics~\cite{Zhang2017,Zhang2018,Wang2018}. Improvements to the quality factor of the optical resonators and the use of a microwave cavity can further reduce the insertion loss and drive voltage required, respectively. For example, increasing optical intrinsic Q to $10^7$~\cite{Zhang2017} will reduce the insertion loss to 0.04 dB, or can reduce both insertion loss and voltage to 0.2 dB and 1 V (see Supplementary Materials). Notably, dynamic control of the shifted light can be achieved by replacing the coupling gap with a microwave-driven Mach-Zehnder interferometer and by applying broadband microwave signals. Moreover, our method for controlling the flow of light in the frequency domain could be applied to other systems, such as mechanics, superconducting qubits, quantum dots, or atomic systems which contain discretized and a continuum of energy levels. The ability to process information in the frequency domain in an efficient, compact, and scalable fashion has the potential to significantly reduce the resource requirements for linear-optical quantum computing~\cite{Lukens2016,Lu2018} and multiplexed quantum communication~\cite{Sinclair2014}. Efficient and on-demand shifting of light may also allow for control of the emission spectrum of solid-state single-photon emitters to create indistinguishable single photons or to produce deterministic single photons from probabilistic emitters~\cite{Joshi2018,GrimauPuigibert2017}. Our reconfigurable frequency shifter could become a fundamental building block for frequency-encoded information processing that offers benefits to telecommunications~\cite{Yoo1996}, radar~\cite{Ghelfi2014}, optical signal processing~\cite{Frankel1998}, spectroscopy~\cite{Demtroder2008}, and laser control applications~\cite{Kohlhaas2012}.

\begin{acknowledgments}
We thank Cheng Wang, Christian Reimer, Joseph Lukens, and Pavel Lougovski for helpful discussion.
This work is supported by the Office of Naval Research (QOMAND N00014-15-1-2761), Air Force Office of Scientific Research (FA9550‐19‐1‐0310), National Science Foundation (ECCS-1839197, ECCS-1541959 and PFI-TT IIP-1827720). Device fabrication was performed at the Harvard University Center for Nanoscale Systems. D.Z. acknowledges support by the Harvard Quantum Initiative. N.S. acknowledges support by the Natural Sciences and Engineering Research Council of Canada (NSERC), the AQT Intelligent Quantum Networks and Technologies (INQNET) research program, and by the DOE/HEP QuantISED program grant, QCCFP (Quantum Communication Channels for Fundamental Physics), award number DE-SC0019219. E. P. acknowledges support by a Draper Fellowship. 
\end{acknowledgments}

%

\pagebreak

\begin{center}
	\textbf{\large Supplemental Materials}
\end{center}
\setcounter{equation}{0}
\setcounter{figure}{0}
\setcounter{table}{0}
\setcounter{page}{1}
\makeatletter
\renewcommand{\theequation}{S\arabic{equation}}
\renewcommand{\thefigure}{S\arabic{figure}}
\setcounter{section}{0}

\section{Device fabrication}
Devices are fabricated from a commercial x-cut lithium niobate (LN) on insulator wafer (NANOLN), with a 600 nm LN layer, 2 $\mathrm{\mu m}$ buried oxide (thermally grown), on a 500 $\mathrm{\mu m}$ Si handle. Electron-beam lithography followed by $\mathrm{Ar^+}$-based reactive ion etching (350 nm etch depth) are used to pattern the optical layer of the device, including the rib waveguides and micro-ring resonators. Microwave electrodes (300 nm of Au) are defined by photolithography followed by electron-beam evaporation and a bilayer lift-off process. The devices are then cladded with 1.6 $\mathrm{\mu m}$-thick $\mathrm{SiO_2}$ using plasma-enhanced chemical evaporation deposition (PECVD). Vias are subsequently patterned using photolithography and etched through the oxide using hydrofluoric acid. Finally, another layer of metal (500 nm of Au) is patterned by photolithography, electron-beam evaporation and lift-off, to form crossovers that connect the bottom electrodes through the vias to ensure the desired polarities for microwave modulation.

\begin{figure*}
\includegraphics{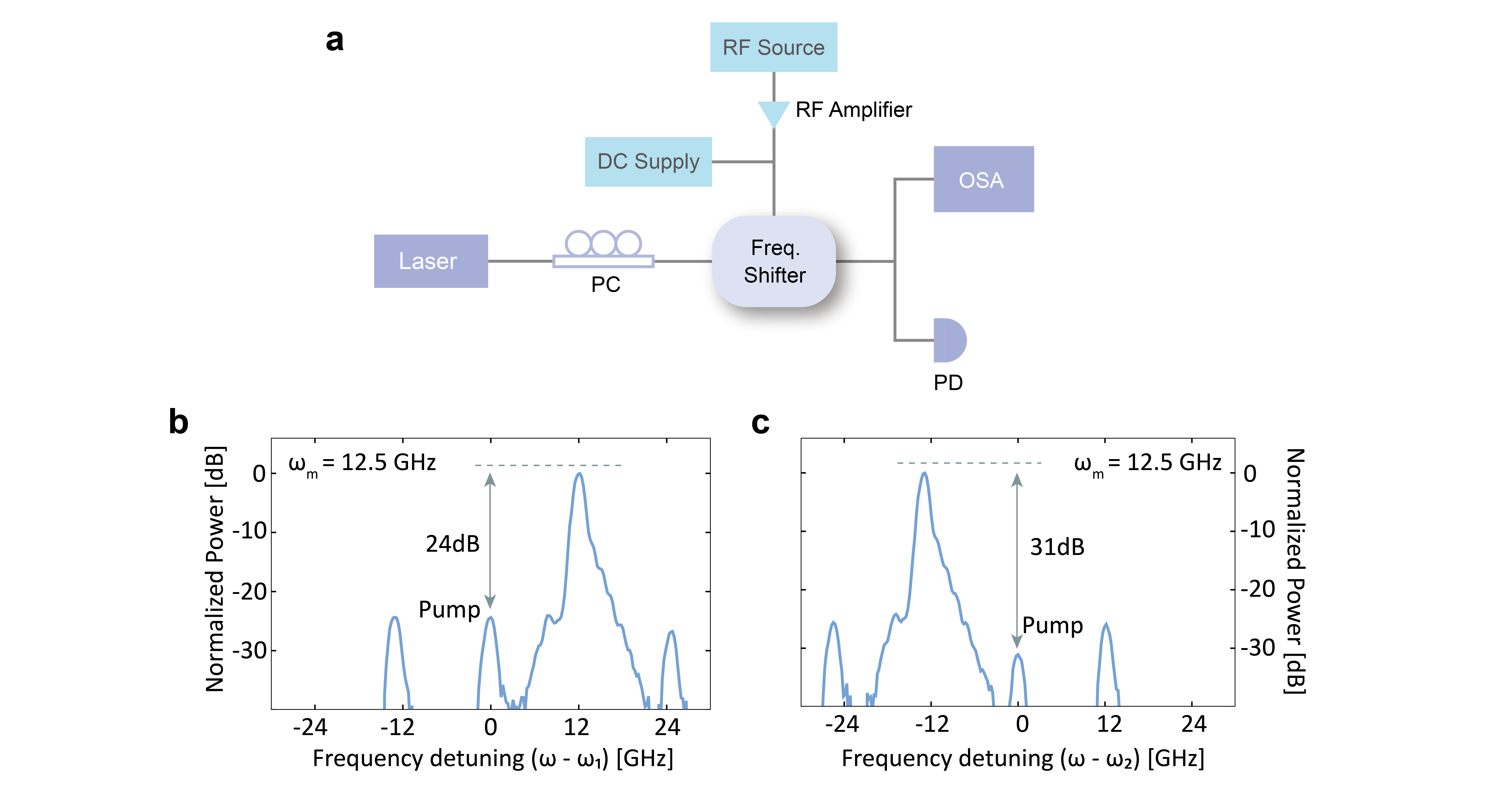}
\caption{\label{Supp1} Setup of frequency shift measurements and optical spectra. a, Set-up for measuring frequency up- and down- shifts. PC, polarization controller; OSA, optical spectrum analyzer; PD, photodetector. b, c, Up- and down- shifts spectra in dB scale from a device with 12.5 GHz doublet splitting. Data corresponds to that in Fig.~\ref{fig2}b and c, respectively, in the main text. }
\end{figure*}

\section{Experimental characterization}
The measurement setup is illustrated in the Fig.~\ref{Supp1}a. Telecommunication-wavelength light from a fiber-coupled tunable laser (SANTEC TSL-510) passes through a polarization controller, and is coupled to the LN chip using a lensed fiber. The output is collimated using an aspheric lens and then sent to an optical spectrum analyzer (OSA) with a spectral resolution of 0.02 nm for characterization of the frequency shift. The microwave signal is generated from a synthesizer followed by a microwave amplifier. After passing through a circulator, the microwave signal is combined with a DC bias through a bias-tee and delivered to the electrodes on the device using an electrical probe. The DC signal tunes the optical resonances of both rings to achieve degenerate condition and form symmetric (S) and anti-symmetric (AS) hybrid modes. The frequency shifts for devices with 12.5 GHz, 11.0 GHz, and 28.2 GHz doublet splitting are measured at pump wavelengths of 1601.3 nm, 1631.5 nm, and 1633.2 nm, respectively. Figure~\ref{Supp1}b and c correspond to the frequency shift shown in Fig.~\ref{fig2}b and c, respectively, on a logarithmic vertical scale. The output spectrum consists of four frequency components: pump frequency, shifted frequency (anti-Stoke/Stoke line), shifted frequency of opposite sign (Stoke/anti-Stoke line), and second-order harmonic frequency.

\begin{figure}
\includegraphics{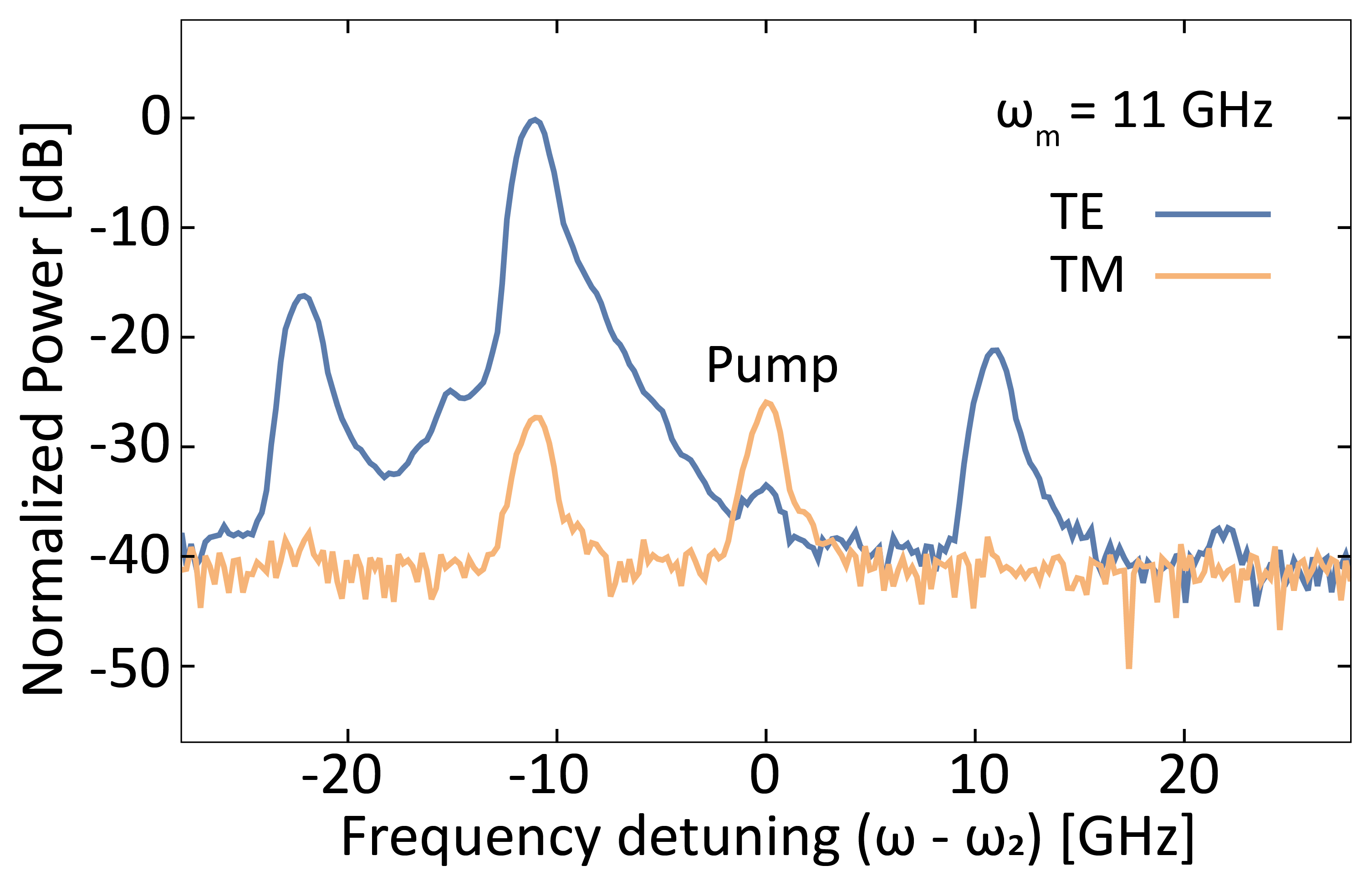}
\caption{\label{Supp2} Polarization of the frequency shifter output. This measurement is performed on a device with 11.0 GHz doublet splitting. The TE and TM components of the output light are measured using an optical spectrum analyzer after passing a polarizer. Here the power is normalized by the summation of TE and TM output powers.}
\end{figure}

The main source of error in estimating the shift efficiency $\eta$ is the imperfect polarization control of the input light. The shifter is designed to operate using the transverse-electric (TE) mode. The control of the purity of the input polarization is limited by the 20 dB -30 dB extinction ratio of the polarization controller. In experiment we verified the polarization of the output light on a device with 11.0 GHz doublet splitting and found the unshifted pump power is predominantly in the transverse-magnetic (TM) mode (see Fig.~\ref{Supp2}). Any TM input will mainly remain unshifted, leading to an underestimation of the optimal shift efficiency.

The experimental demonstration of the swap operation is performed at a wavelength of 1560.6 nm (setup shown in Fig.~\ref{fig3}a). We first set the frequency of two laser beams to be far detuned from the doublet resonance and measure the time-domain audio signals as references. This corresponds to the case in which the signals are not swapped. We then tune the frequency of each laser beam to be on resonance with one of the modes of the doublet, i.e. laser 1 (2) in S (AS) mode. In this case, frequency components around laser beam 1 are up-shifted and components around the frequency of laser beam 2 are down-shifted. The amplitudes of the time domain signal before and after swapping are renormalized for comparison in Fig.~\ref{fig3}b.

\section{Device Parameter}
Our device parameters are characterized by sweeping the laser wavelength from 1580 nm to 1680 nm and measuring the transmission spectrum of the devices. Four parameters are extracted for each device: waveguide-cavity coupling $\gamma$, intrinsic loss rate $\kappa_i$ of the S and AS modes, linewidth $\kappa=\gamma/2+\kappa_i$ of the S and AS modes, and intrinsic quality factor $Q_\mathrm{intrinsic}$ of the S and AS modes.

For the device with 12.5 GHz doublet splitting, we calculate $\gamma=2\pi \times3.95$ GHz, $\kappa_i=2\pi\times0.28$ GHz, $\kappa=2\pi\times1.97$ GHz, and $Q_\mathrm{intrinsic} \sim0.65\times10^6$ for the doublet at 1601.3 nm. For the 11.0 GHz device, we selected the resonances at 1631.5 nm and calculate $\gamma=2\pi\times4.98$ GHz, $\kappa_i=2\pi\times0.33$ GHz, $\kappa=2\pi\times2.82$ GHz, and $Q_\mathrm{intrinsic} \sim 0.56\times10^6$. For the 28.2 GHz shift device, we analyze the resonances at 1633.2 nm and get $\gamma=2\pi\times5.31$ GHz, $\kappa_i=2\pi\times0.17$ GHz, $\kappa=2\pi\times2.82$ GHz, and $Q_\mathrm{intrinsic}\sim 1.1\times10^6$. 

\section{Characterization and limitation of insertion loss}
The insertion loss for the devices having doublet splitting of 12.5 GHz, 11.0 GHz, and 28.2 GHz are measured to be 1.22 dB, 1.25 dB, and 0.45 dB, respectively. The device insertion loss IL is defined to be the loss experienced by the light that travels through a device. Since our device is resonance-based, light will not go through the device if the laser is far detuned from resonance. Therefore, $IL$ is determined by comparing the transmission when the laser is tuned on and far off resonance. The main source of error is the Fabry-Perot fringes that are induced by the two facets of the chip. These fringes produce a variation of the off-resonance transmission. We address this uncertainty by averaging multiple off-resonance power at different wavelengths.

\begin{figure}
\includegraphics{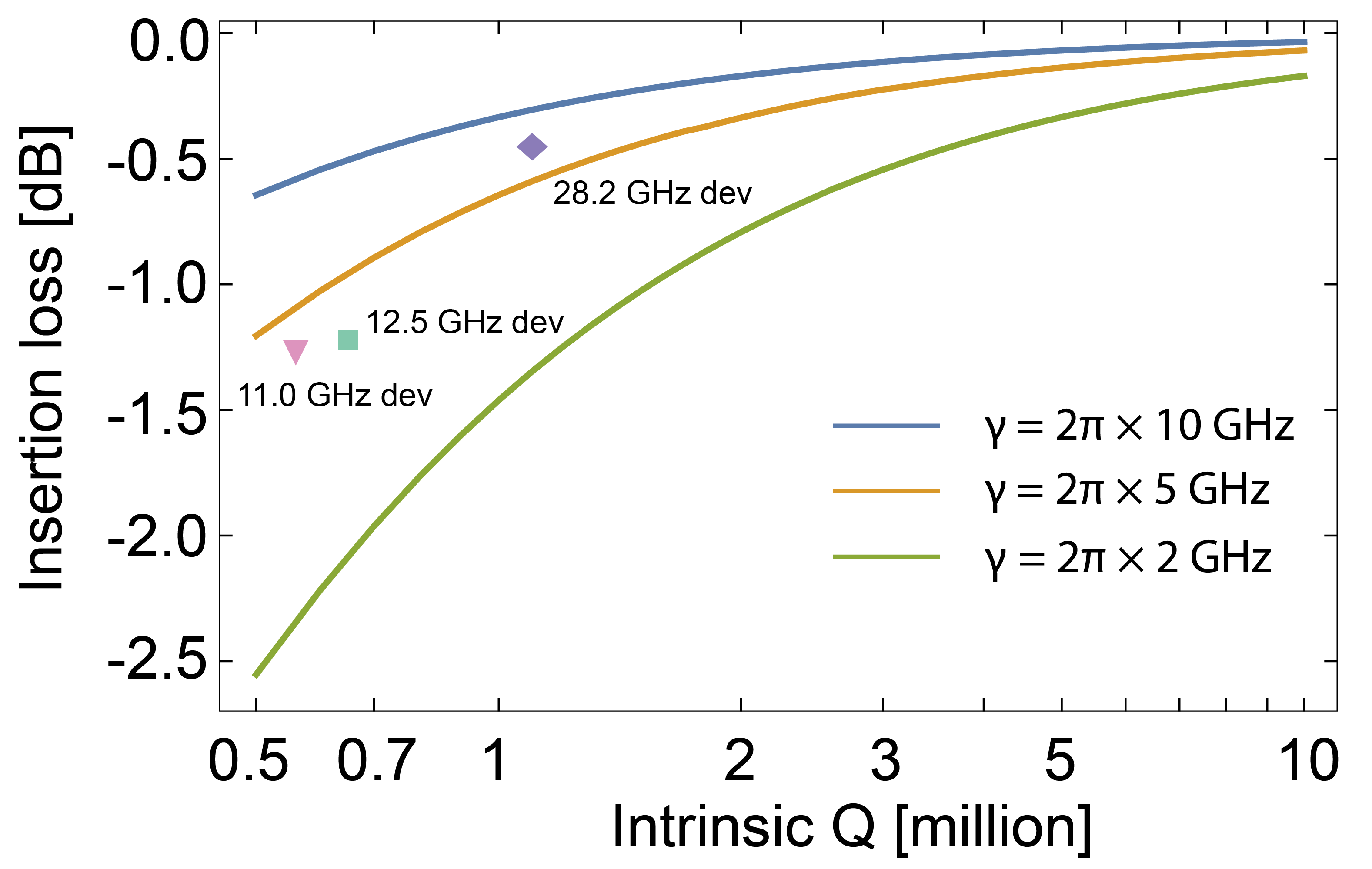}
\caption{\label{Supp3} Ultimate limitation of the device insertion loss. Predicted device insertion loss with varied $Q_\mathrm{intrinsic}$ for different waveguide-ring couplings $\gamma$. The theoretical curve of insertion loss is calculated under the optimal condition of both pump and microwave detunings are zero. The measured insertion loss of the 11.0 GHz, 12.5 GHz, and 28.2 GHz devices used in this work are labeled with a triangle, square, and diamond, respectively.}
\end{figure}

To study the ultimate limit of the device insertion loss, we theoretically calculate IL as a function of $Q_\mathrm{intrinsic}$ (Fig.~\ref{Supp3}) with different waveguide-ring coupling rates $\gamma$. Since our device operates in the strongly over-coupled regime (loaded quality factor $Q_\mathrm{load} \sim 80,000$ for S and AS modes), light only takes a few roundtrips inside the cavity with small propagation losses. As a result, the device insertion loss is close to that of a short bare waveguide. Thus increasing the ratio between waveguide-ring coupling rate $\gamma$ and the intrinsic loss rate $\kappa_i$ can reduce the device insertion loss, as shown by Fig.~\ref{Supp3}. For example, increasing $Q_\mathrm{intrinsic}$ to $10^7$~\cite{Zhang2017} will reduce the insertion loss to 0.04 dB for $\gamma\sim 2\pi\times 8.6$ GHz. 

Moreover, although decreasing $\gamma$ leads to a larger insertion loss, the required voltage can be largely reduced. For example, for $Q_\mathrm{intrinsic} \sim 10^7$, reducing $\gamma$ to $2\pi\times3.5$ GHz gives $IL=0.1$ dB and requires only 2 V microwave peak-voltage for a frequency shift of 28.2 GHz. Further reducing $\gamma$ to $2\pi\times1.7$ GHz yields $IL=0.2$ dB and requires only 1 V microwave peak-voltage. The current device with a 28.2 GHz doublet splitting requires on-chip peak-voltage 3.1 V (see next section Ultimate limit of the magnitude of frequency shift). 

\section{Ultimate limit of the magnitude of frequency shift}
This limit is due to the electrical circuit and not the optical components. In the main text we showed that our device efficiency and insertion loss are independent of the magnitude of frequency shift. To increase the magnitude of frequency shift, a smaller coupling gap between the two cavities can be used to increase doublet splitting. Since our cavity has a free spectral range of 250 GHz, it is not restricting the doublet splittings of our current device. Also, the intrinsic loss $\kappa_i$ and waveguide-ring coupling $\gamma$ can be maintained when the doublet splitting is increased. However, the electrode performance will degrade at high microwave frequencies. The current electrode is designed to be a capacitor to induce an electric field across the lithium niobate cavity. The impedance of the capacitor decreases with increasing microwave frequency. Thus, the circuit ultimately becomes a short load instead of an open load at high frequencies, and the voltage delivered to the device is reduced.

\begin{figure*}
\includegraphics{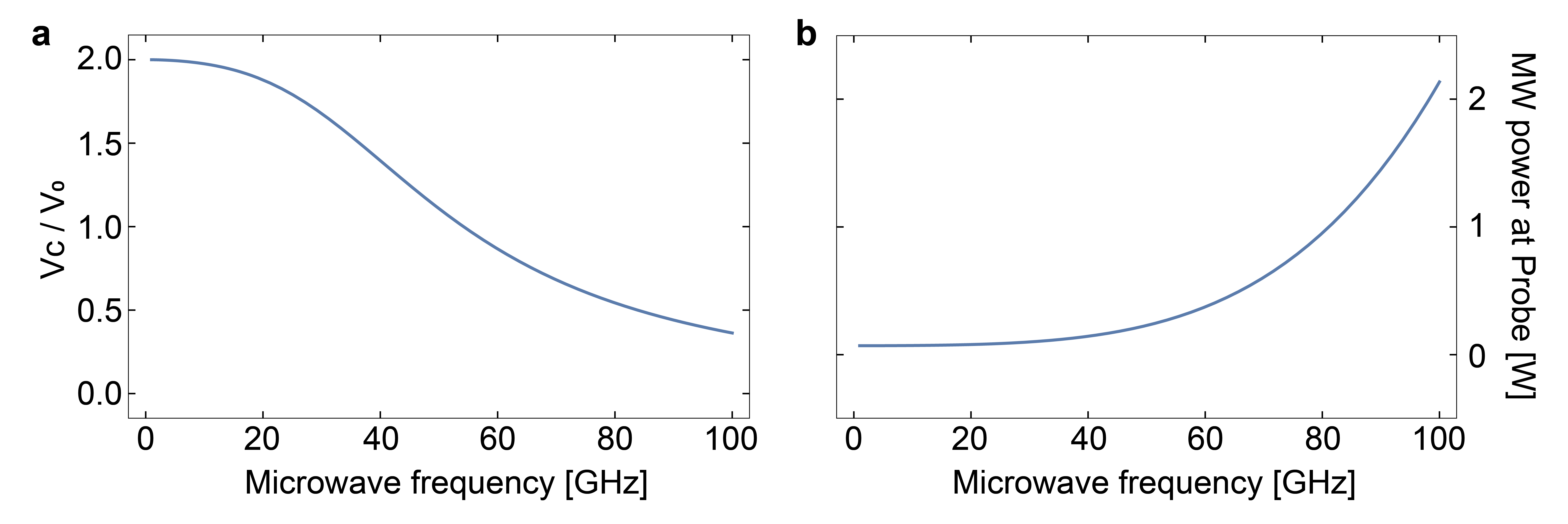}
\caption{\label{Supp4} Simulated electrode performance for varied microwave frequencies. a, The relative voltage delivered to the capacitor $V_c$ and probe $V_0$ for varied microwave frequencies. b, Power required at the probe (50 $\Omega$) for a maximal shift efficiency as a function of microwave frequency. Our current 28.2 GHz device requires 96 mW microwave power (3.1 V microwave peak-voltage). We assume that the voltage $V_c$ delivered to the capacitor is kept same as the peak-voltage on capacitor ($V_c=5.3$ V) of our 28.2 GHz device when reaching 98.7\% efficiency. MW, microwave.}
\end{figure*}

To quantitatively estimate this frequency limitation from the electrode, we simulate our electrode and obtain a 0.11 pF capacitance with a 0.12 nH inductance and 1.5 $\Omega$ resistance. We then use an LCR model to calculate the voltage delivered to the capacitor. Figure~\ref{Supp4}a shows the voltage $V_c$ on the capacitor divided by the voltage $V_0$ on the 50 $\Omega$ input probe as a function of microwave frequency. For example, it is obtained that $V_c=1.72V_0$ at 28.2 GHz and $V_c=0.5V_0$ at 84.0 GHz, indicating a lower voltage delivery from the 50 $\Omega$ input probe to the capacitor. To estimate the power required at high frequencies, we note the required coherent coupling strength is $\Omega \sim 2.65$ GHz to achieve the shift efficiency 98.7\% using the device with 28.2 GHz doublet splitting. Combined with the electro-optic coefficient ($\sim0.5$ GHz/V ~\cite{Zhang2018}), this coherent coupling strength $\Omega$ corresponds to a required voltage of $V_c  =5.3$ V on the capacitor and $V_0=3.1$ V on the 50 $\Omega$ input probe. In Fig.~\ref{Supp4}b we plot the microwave input power that is needed from the probe as a function of frequency in order to maintain $V_c=5.3$ V. It can be seen that the required power becomes dramatically higher ($> 1$ W) above ~84 GHz to compensate for the inefficient capacitive drive. To circumvent this problem, one could either design an electrode with lower capacitance or use a taper between the probe and the electrode which transforms the transmission line impedance to lower values and increases the effective RC frequency limit.

\section{Numerical simulation}
To numerically simulate the device, we use a system of phase-modulated coupled cavities to demonstrate an efficient frequency shift. The Hamiltonian of this system can be described as 

\begin{equation}
\begin{aligned}
    H= &\ \omega_\mathrm{ring1} a_1^\dagger a_1+\omega_\mathrm{ring2} a_2^\dagger a_2+\mu(a_1^\dagger a_2+a_1 a_2^\dagger)\\
    &+\Omega\cos (\omega_mt+\phi) (a_1^\dagger a_1-a_2^\dagger a_2 )
\end{aligned}
\end{equation}
where $a_1$ and $a_2$ are the annihilation operators of the optical fields in cavities 1 and 2, while $\omega_\mathrm{ring1}$ and $\omega_\mathrm{ring2}$ are the corresponding cavity resonance frequencies, $\mu$ is the coupling strength between the two optical cavities due to the evanescent coupling, $\Omega$ is the modulation strength which is proportional to the microwave peak-voltage, $\omega_m$ is the microwave frequency, and $\phi$ is the phase of the microwave signal. Here and henceforth $\hbar=1$. The minus sign in the last term is due to the modulation configuration of our system, in which the applied external voltage increases (decrease) the frequency of the resonance of ring 1 (2). 

The equations of motion are based on the Heisenberg-Langevin equation derived from the Hamiltonian of the system. Accordingly, we obtain the equation of motion in the laser rotating frame:
\begin{equation}
\begin{aligned}
    \dot a_1= & \ (-i(\omega_\mathrm{ring1}-\omega_L)-\frac{\gamma+\kappa_i}{2}) a_1\\
    & -i\Omega \cos(\omega_mt+\phi)a_1-i\mu a_2-\sqrt\gamma\alpha_\mathrm{in}\\
    \dot a_2= & \  (-i(\omega_\mathrm{ring2}-\omega_L)-\frac{\kappa_i}{2}) a_2\\
    & +i\Omega \cos(\omega_mt+\phi)a_2-i\mu a_1
\end{aligned}
\end{equation}
where $\gamma$ is the waveguide-ring coupling and $\kappa_i$ is the intrinsic loss of the ring. The amplitude of the input field is $\alpha_\mathrm{in}=\sqrt{P_\mathrm{in}/\omega_L}$, and the input pump power and frequency are $P_\mathrm{in}$ and $\omega_L$, respectively.  The simulation is performed by numerically solving these equations of motion to obtain the output field amplitude as $a_\mathrm{out}=\alpha_\mathrm{in}+\sqrt{\gamma}a_1$. In the end, a Fourier transform is performed to analyze the frequency component of the output field. 

\section{Theoretical analysis}
The above equations can be solved analytically using several transformations and approximations. We consider the case that the two rings are identical, i.e. $\omega_\mathrm{ring1}=\omega_\mathrm{ring2}=\omega_0$. Then the Hamiltonian can be transformed to the basis of symmetric (S) mode $c_1=\frac{1}{\sqrt 2}(a_1+a_2)$ and anti-symmetric (AS) mode $c_2=\frac{1}{\sqrt 2}(a_1-a_2)$. In this basis, the Hamiltonian of the system transforms to
\begin{equation}
    H=\omega_1 c_1^\dagger c_1+\omega_2 c_2^\dagger c_2+\Omega \cos (\omega_mt+\phi) (c_1^\dagger c_2+h.c.)
\end{equation}
where $\omega_1=\omega_0-\mu$ and $\omega_2=\omega_0+\mu$. As before, the equations of motion in the basic of $c_1$, $c_2$ can be derived from the Heisenberg-Langevin equations, yielding:
\begin{equation}
\begin{aligned}
    & \dot c_1=(-i\omega_1-\frac{\kappa_1}{2}) c_1-i\Omega \cos(\omega_mt+\phi)c_2 -\sqrt{\kappa_{e1}}\alpha_\mathrm{in}e^{-i\omega_L t}\\
    & \dot c_2=(-i\omega_2-\frac{\kappa_2}{2}) c_2-i\Omega \cos(\omega_mt+\phi)c_1 -\sqrt{\kappa_{e2}}\alpha_\mathrm{in}e^{-i\omega_L t}
\end{aligned}
\end{equation}
where $\kappa_{e1}$ and $\kappa_{e2}$ are the external loss rate to the waveguide for the S and AS modes, $\kappa_i$ is the intrinsic loss, $\kappa_j=\kappa_{e_j}+\kappa_i$ is the total linewidth of S ($j=1$) and AS ($j=2$) mode, $\omega_L$ and $\alpha_\mathrm{in}$ is the frequency and amplitude of the input field. This formalism is also consistent with the formalism in Wade et al~\cite{Wade2015} and Zhang et al~\cite{Zhang2018}. For our system $\kappa_{e1}=\kappa_{e2}=\gamma/2$ because the S and AS modes physically occupy both rings, while only ring1 ($a_1$) is coupled to the bus waveguide with a rate $\gamma$. For simplicity, we assume $\kappa_e\equiv\kappa_{e1}=\kappa_{e2}$ and $\kappa\equiv\kappa_1=\kappa_2$.

With the symmetric mode being pumped, the modes $c_1$,$c_2$ can be replaced by their slowly varying amplitudes as $c_1\rightarrow c_1 e^{-i\omega_L t}$, $c_2\rightarrow c_2 e^{-i\omega_L t} e^{-i\omega_m t}$. Under the rotating wave approximation, the equations of motion become 
\begin{equation}
\begin{aligned}
    & \dot c_1=(i\Delta-\frac{\kappa}{2}) c_1-i\frac{\Omega}{2} e^{i\phi} c_2 -\sqrt{\kappa_{e}}\alpha_\mathrm{in}\\
    & \dot c_2=(i\Delta+i\delta-\frac{\kappa}{2}) c_2-i\frac{\Omega}{2} e^{-i\phi}c_1 -\sqrt{\kappa_{e}}\alpha_\mathrm{in}e^{i\omega_m t}
\end{aligned}
\end{equation}
where $\Delta=\omega_L-\omega_1$ is the laser detuning with respect to the S mode and $\delta=\omega_m-2\mu$ is the detuning of microwave field with respect to the S-AS doublet splitting. Using the fact that $c_2$ is off resonantly pumped, the pump term in the equation of motion of $c_2$ can be neglected, and the steady-state solution can be obtained (for simplicity we set $\phi=0$):
\begin{equation}
\begin{aligned}
    c_1= & \ 
    \frac{\sqrt{\kappa_e} \alpha_\mathrm{in}}{
    i\Delta-\frac \kappa 2
    +\frac{\Omega^4/4}{i\Delta+i\delta-\frac \kappa 2}}\\
    c_2= & \ 
    i \frac \Omega 2 
    \frac{\sqrt{\kappa_e} \alpha_\mathrm{in}}{
    (i\Delta-\frac \kappa 2)(i\Delta+i\delta-\frac \kappa 2)+\frac{\Omega^2}{4}}
\end{aligned}
\end{equation}
The output field will then be 
\begin{equation}
    a_\mathrm{out}=\alpha_\mathrm{in} e^{-i\omega_L t}
    +\sqrt \kappa_e (c_1 e^{-i\omega_L t}+c_2 e^{-i\omega_L t} e^{-i\omega_m t})
\end{equation}

The output field can be rewritten as two different frequency components $a_\mathrm{out}=A_0 e^{-i\omega_L t}+A_+ e^{-i\omega_L t} e^{-i\omega_m t}$ with $A_0=\alpha_\mathrm{in}+\sqrt{\kappa_{e}}c_1$ and $A_+=\sqrt{\kappa_e}c_2$. For the case of zero optical and microwave detuning ($\Delta=0$ and $\delta=0$), the pump component $A_0$ becomes $\alpha_\mathrm{in}\left(1+\frac{\kappa_e}{-\frac \kappa 2 +\frac{\Omega^2/4}{-\frac \kappa 2}}\right)$, which indicates that the coherent coupling $\Omega$ introduces an effective intrinsic loss channel for mode $c_1$ as expected. The total loss for mode $c_1$ can be written as $\kappa_{1_\mathrm{eff}}=\kappa\left(1+(\frac \Omega \kappa)^2 \right)$, where the factor $\Omega^2/\kappa^2$ is the loss rate that is induced by coupling to another mode and plays a similar role as the Purcell effect in cavity quantum electrodynamics. This effective loss balances the large external loss channel ($\kappa_e$) to the waveguide and leads to the complete suppression of the pump. Considering the case $\kappa_e\gg\kappa_i$, we obtain a simple condition for generating a unidirectional frequency shift: $\kappa_e=\Omega^2/\kappa_e$, which means $\Omega=\kappa_e$ when the intrinsic loss is negligible (as is the case for our device). 

The concept of balancing the coupling rates can be intuitively understood using impedance matching. For example, when an optical cavity with an intrinsic loss rate $\kappa_i$ is coupled to a waveguide with a coupling strength $\kappa_e$, some of the light is coupled back to the waveguide. This is identical to the microwave reflection coefficient in transmission-line theory with effective reflection coefficient $\Gamma=\frac{\kappa_i-\kappa_e}{\kappa_i+\kappa_e}$. In this picture, the case of strong under-coupling ($\kappa_i\gg\kappa_e$) and over-coupling ($\kappa_i\ll\kappa_e$) between a waveguide and a cavity corresponds to open and short circuits, respectively. This interpretation helps to understand the cascaded frequency shifting scheme that is discussed in Fig.~\ref{fig4} in the main text, in which photons propagate through a ladder of energy levels without reflection to realize unidirectional frequency shifts of $>$100 GHz.

Moreover, the current microwave bandwidth of our device is at least 3 GHz, benefitting from the strong over-coupling of the S and AS modes (linewidths $\sim 2\pi\times2.8$ GHz) to the waveguide. We note that the microwave frequency can exceed the linewidth of the S and AS modes due to the power broadening of the modes by the strong coherent coupling $\Omega$. This effect could be understood by considering the total loss rate $\kappa_{1_\mathrm{eff}}=\kappa\left(1+\left(\frac \Omega \kappa\right)^2\right)$ of mode $c_1$, in which the strong $\Omega$ leads to a larger loss rate of $c_1$ thus a broader effective linewidth. A similar analysis can be performed on the AS mode $c_2$.

\section{Unitary transformation of the frequency beam splitter}
Here we show that our frequency beam splitter acts on discrete frequency modes according to the same unitary transformation that a beam splitter obeys when acting on spatial modes (paths). Considering the case where $\kappa_i\rightarrow 0$, i.e. that with vanishing insertion loss, pumping of the S mode with $\Delta=0$ and $\delta=0$ gives
\begin{equation}
    A_0=\frac{\Omega^2-\kappa_e^2}{\Omega^2+\kappa_e^2}     \quad        
    A_+=ie^{-i\phi}  \frac{2\Omega \kappa_e}{\Omega^2+\kappa_e^2}
\end{equation}

where $A_0$ and $A_+$ are the two frequency components of the output field $a_\mathrm{out}=A_0 e^{-i\omega_L t}+A_+ e^{-i\omega_L t} e^{-i\omega_m t}$ that are discussed before and are normalized by the input field $\alpha_\mathrm{in}$, alternatively $\alpha_\mathrm{in}=1$. Based on this result, we set $A_0=\cos \theta$ and $A_+=ie^{-i\phi}\sin \theta$ with $\theta=\theta(\Omega)$. Similar results in the case of pumping the AS mode can be obtained. Finally, the operator of our frequency splitter is expressed as:

\begin{equation}
U=\left(
\begin{array}{cc}
     \cos \theta(\Omega) &  ie^{-i\phi} \sin \theta(\Omega) \\
     ie^{i\phi} \sin \theta(\Omega) &  \cos \theta(\Omega)
\end{array}
\right)
\end{equation}
where the splitting ratio is controlled by $\Omega$, which is governed by the microwave power, while the phase is tuned by the microwave phase $\phi$. We note that our EO device is coherent and does not change other degrees of freedom (e.g. polarization or spatial modes) or introduce noise photons from the microwave drive, which is of importance for quantum optics and information tasks (e.g. quantum communication and computing).

\section{Simulation of cascaded frequency shift}
The numerical simulation is performed based on the equations of motion for the system of cascaded frequency shift derived by a similar approach (Heisenberg-Langevin equation) that are discussed above. In the simulation, we assume all three cavities to have an intrinsic $Q \sim 1.8\times10^6$, which corresponds to an intrinsic loss rate $\kappa_i=2\pi\times100$ MHz. The waveguide-mode coupling of ring 1 is $\kappa_e=2\pi\times3$ GHz, the ring 1-ring 2 evanescent coupling is $\mu_1=2\pi\times1.5$ GHz, and the ring 2-ring 3 coupling is $\mu_2=2\pi\times3$ GHz. The coupling induced by microwave modulation is $\Omega=2\pi\times3$ GHz, which is similar to the value that we used in our current devices.
\newpage

\end{document}